\documentclass[sigconf,nonacm]{acmart}
\setcopyright{none}
\renewcommand\shortauthors{}
\usepackage{kotex}
\usepackage{placeins}
\usepackage{float}
\usepackage{microtype}
\usepackage{enumitem}
\usepackage{booktabs}
\usepackage{array}
\usepackage{multirow}
\usepackage{tabularx}
\usepackage{subcaption}
\usepackage{balance}
\usepackage{hyperref}
\usepackage{pgfplots}
\usetikzlibrary{patterns,patterns.meta}
\pgfplotsset{compat=1.18}
\usepackage[normalem]{ulem}
\usepackage{pgfplotstable}
\usepackage{mathtools}
\newcolumntype{C}{>{\centering\arraybackslash}p{2cm}}
\hypersetup{
    colorlinks=true,
    linkcolor=blue,
    urlcolor=cyan
}
\usepackage{soul}

\usepackage{tikz}
\usepackage{pgfplots}
\pgfplotsset{compat=1.18}
\usepackage{caption}


\definecolor{darkblue}{rgb}{0.0, 0.0, 0.55}
\definecolor{ao(english)}{rgb}{0.0, 0.5, 0.0}
\definecolor{coolblack}{rgb}{0.0, 0.18, 0.39}
\definecolor{purpleheart}{rgb}{0.41, 0.21, 0.61}
\definecolor{pastelviolet}{rgb}{0.8, 0.6, 0.79}
\definecolor{lightskyblue}{rgb}{0.53, 0.81, 0.98}
\definecolor{palecornflowerblue}{rgb}{0.67, 0.8, 0.94}
\definecolor{lightmauve}{rgb}{0.86, 0.82, 1.0}
\definecolor{lightpastelpurple}{rgb}{0.69, 0.61, 0.85}
\definecolor{bronze}{rgb}{0.8, 0.5, 0.2}
\definecolor{armygreen}{rgb}{0.29, 0.33, 0.13}
\definecolor{darkpowderblue}{rgb}{0.0, 0.2, 0.6}
\definecolor{falured}{rgb}{0.5, 0.09, 0.09}
\definecolor{outerspace}{rgb}{0.25, 0.29, 0.3}
\definecolor{tangerine}{rgb}{0.95, 0.52, 0.0}
\definecolor{seagreen}{rgb}{0.18, 0.55, 0.34}
\definecolor{springgreen}{rgb}{0.0, 1.0, 0.5}
\definecolor{applegreen}{rgb}{0.55,0.71,0.0}
\definecolor{amethyst}{rgb}{0.6,0.4,0.8}
\definecolor{amber}{rgb}{1.0,0.49,0.0}
\definecolor{darkgreen}{rgb}{0,0.4,0}

\usepackage[table]{xcolor}
\usepackage{pgfplots}
\usetikzlibrary{plotmarks}
\usepgfplotslibrary{groupplots}

\AtBeginDocument{%
  }

\begin{document}

\title{Embedding-aware Polarization Management in Signed Networks}

\author{Jeonghan Son}
\affiliation{
  \institution{Ulsan National Institute of Science and Technology (UNIST)}
  \city{Ulsan}
  \country{South Korea}
}
\email{sjh000606@unist.ac.kr}

\author{Kyungsik Han}
\affiliation{
  \institution{Hanyang University}
  \city{Seoul}
  \country{South Korea}
}
\email{kyungsikhan@hanyang.ac.kr}

\author{Yeon-Chang Lee}
\authornote{Corresponding author.}
\affiliation{
  \institution{Ulsan National Institute of Science and Technology (UNIST)}
  \city{Ulsan}
  \country{South Korea}
}
\email{yeonchang@unist.ac.kr}

\begin{abstract}
Signed network embeddings (SNE) are widely used to represent networks with positive and negative relations, but their repeated use in downstream analysis pipelines can inadvertently reinforce structural polarization. 
Existing polarization measures are largely designed for \textit{unsigned} networks or rely on \textit{predefined} opinion states, limiting their applicability to embedding-based analysis in signed settings. 
We propose \ours, a unified polarization management framework that jointly measures and mitigates polarization in the embedding space. 
\ours\ introduces an embedding-based polarization measure grounded in effective resistance and a structure-aware mitigation strategy via localized augmentation through structurally balanced intermediary nodes. 
Experiments on real-world signed networks demonstrate that \ours\ effectively mitigates polarization while preserving task-relevant network structure.
The codebase of \ours\ is currently available at
\url{https://github.com/JeonghanSon/EPM-Embedding-aware-Polarization-Management}.

\end{abstract}

\newcommand{\spec}{{\it spec.}}
\newcommand{\aka}{{\it a.k.a.}}
\newcommand{\ie}{{\it i.e.}}
\newcommand{\eg}{{\it e.g.}}
\newcommand{\ours}{\textsc{\textsf{EPM}}}
\newcommand{\blue}{\textcolor{blue}}

\newcommand{\mj}[1]{\textcolor{blue}{[MJ: #1]}}
\newcommand{\jw}[1]{\textcolor{green}{[JW: #1]}}
\newcommand{\yc}[1]{\textcolor{red}{[YC: #1]}}

\newcommand{\oursall}{\textsc{\textsf{{TraceRec(all)}}}}
\newcommand{\oursforward}{\textsc{\textsf{{TraceRec(forward)}}}}
\newcommand{\oursrecent}{\textsc{\textsf{{TraceRec(recent)}}}}

\newcommand{\ourswoproj}{\textsc{\textsf{{TraceRec(w/o proj)}}}}

\maketitle
\pagestyle{plain}
\section{Introduction} \label{sec:intro} \noindent\textbf{Background.}
Real-world social interactions often involve relations with opposite meanings, such as trust versus distrust or agreement versus disagreement \cite{heider_jpsych46_1,cartwright_psychrev56_1}.
Signed networks provide a standard representation for modeling such dual relations by labeling edges as positive or negative \cite{leskovec_chi10_1}.
These networks have been widely used across diverse domains, including online social platforms \cite{leskovec_chi10_1}, recommender systems \cite{guha_www04_1,massa_recsys07_1}, and reputation systems \cite{kunegis_www09_1,leskovec_chi10_1}. 

To represent such signed networks, extensive research has focused on \textbf{signed network embeddings (SNE)} \cite{yuan_sne_pakdd17,kim_side_www18,kim_trustsgcn_sigir23,lee_asine_sigir20,li_signedgat_aaai20}, which learn low-dimensional vector representations of nodes while preserving positive and negative relations. 
By incorporating social theories such as balance theory \cite{derr_sgcn_icdm18,huang_sigat_icann19,huang_sdgnn_aaai21}, these methods capture key signed structural patterns and achieve strong performance in downstream tasks such as signed link prediction \cite{yang_friendorfrenemy_sigir12,xu_signedlatent_kdd19,fiorini_sigmanet_aaai23,huang_sdgnn_aaai21} and personalized ranking \cite{jung2016personalized,jung2020random,lee2021look}.
These tasks support real-world applications such as recommendation in review and reputation systems \cite{kunegis_www09_1,lee_asine_sigir20,kim_trustsgcn_sigir23} and have also been applied to the analysis and prediction of international conflicts~\cite{MorrisonKG23,DiazDiazBE24,fritz2025}.



\vspace{+1mm}
\noindent\textbf{Motivation.}
As SNE are increasingly adopted in real-world systems, they are typically embedded within broader decision-making pipelines \cite{derr_sgcn_icdm18,kim_polardsn_cikm24} rather than used as standalone analysis tools. 
In practice, learned embeddings are repeatedly reused to infer signed links, guide recommendation or ranking decisions, and update network structures for subsequent model retraining. 

While such reuse is natural and often necessary, it may introduce unintended long-term effects. 
When embeddings that accurately encode positive and negative relations are repeatedly deployed, interactions between opposing groups may gradually diminish.
Over time, this process can reinforce existing structural divisions, contributing to increased \textbf{polarization} in the signed network \cite{cinelli_echo_pnas21,santos_linkrec_pnas21}.
Importantly, this phenomenon does not stem from inaccuracies in individual embedding models; rather, it emerges from the repeated deployment of accurate representations within dynamic systems.
These observations raise a fundamental question of \textit{how does the repeated deployment of accurate SNE influence polarization and broader structural outcomes at the ecosystem level}. 



\begin{figure}[t]
    \centering
    \captionsetup{type=figure,skip=8pt}
    \includegraphics[width=0.9\linewidth]{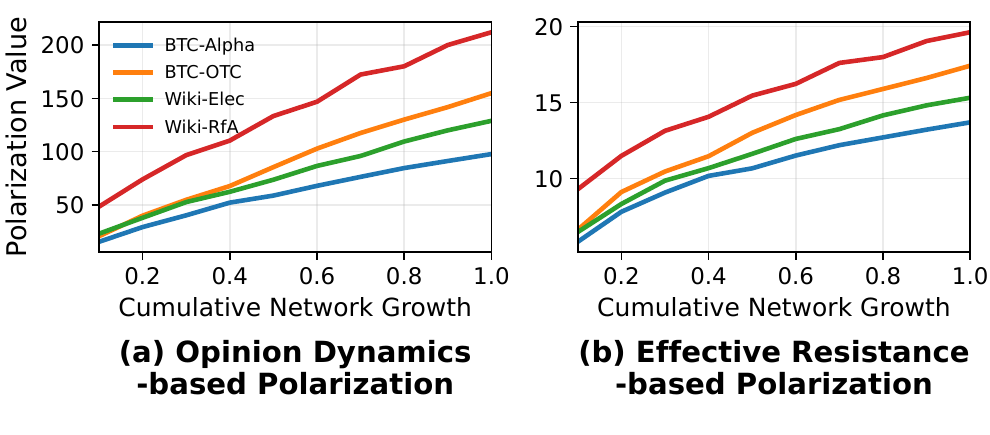}
    \vspace{-0.2cm}
    \caption{Temporal evolution of structural polarization in signed networks.
    Across all datasets, structural polarization consistently increases as networks grow, suggesting that polarization can accumulate as signed networks evolve.
    }
    \label{fig:intro}
    \vspace{-0.55cm}
\end{figure}

\vspace{1mm}
\noindent\textbf{Polarization Measurement.}
Answering this question requires a principled way to quantify polarization in signed networks.
Related ideas have been explored in prior work, spanning multiple perspectives on polarization.
Early work \cite{dandekar_pnas13_1,musco_www18_1,friedkin_johnsen_opinion90} considered \textit{opinion dynamics}, modeling node states as explicit opinions that evolve according to predefined update rules.
Other studies \cite{garimella_wsdm17_1,garimella_tsc18_1,guerra_polarization_icwsm13,coletto_controversy_osnem17,emamgholizadeh_brw_snam20,cota_echo_epjds19} examined \textit{interaction patterns}, using random walks or information flow to quantify separation between groups.
More recently, graph-theoretic approaches \cite{hohmann_sciadv23_1} have proposed \textit{structural measures} of polarization, including effective resistance or generalized distances.

To examine the question raised earlier, we conduct a preliminary empirical analysis on four real-world temporal signed network datasets using two polarization measures based on opinion dynamics \cite{musco_www18_1} and effective resistance \cite{hohmann_sciadv23_1}, respectively.
For each dataset, we construct a sequence of static network snapshots by cumulatively aggregating signed edges over time and compute polarization values for each snapshot.
As shown in Figure~\ref{fig:intro}, polarization consistently increases as networks grow across all datasets, indicating that structural polarization naturally intensifies as signed networks evolve over time.
This observation raises an important concern that accurately modeling such evolving signed structures may inadvertently reinforce polarization when the learned representations are repeatedly reused in downstream systems. This possibility motivates the central research question of this work.

\vspace{1mm}
\noindent\textbf{Limitations.}
While existing polarization measures are useful for analyzing structural division in networks, they are not designed to address the question posed in this work.
Most measures \cite{dandekar_pnas13_1,musco_www18_1,garimella_tsc18_1,hohmann_sciadv23_1,garcia_polarization_policyinternet15}
 are developed under \textit{unsigned} network assumptions and treat polarization as a descriptive property of network topology. 

Such assumptions become restrictive in \textit{signed} networks, where positive and negative relations play fundamentally different roles in shaping group structure.
However, many existing measures do not explicitly account for how these relations interact or evolve.
Moreover, existing measures are largely decoupled from representation learning: node states are often assumed to be \textit{scalar-valued opinions}, making it unclear how antagonistic structures encoded in the embedding space learned by SNE should be reflected in polarization measurement.
In addition, while some prior studies \cite{musco_www18_1,garimella_wsdm17_1,haddadan2021repbublik}
 consider polarization \textit{mitigation} alongside measurement, most mitigation approaches focus on directly
reducing polarization scores, often through \textit{global network restructuring}.
While such approaches may effectively lower measured polarization, they often overlook downstream task utility, limiting their suitability for real-world applications where preserving task-relevant structure is essential.

\vspace{1mm}
\noindent\textbf{Our Work.}
To address the limitations discussed above, we propose \textbf{\ours} (\textbf{E}mbedding-aware \textbf{P}olarization \textbf{M}anagement), a unified framework for \textit{measuring} and \textit{mitigating} polarization in signed networks. 
This design enables polarization to be assessed and reduced in a manner consistent with the embedding space used in practice.


On the \textbf{measurement side}, \ours\ introduces an embedding-aware polarization metric, $P_{\mathcal{G},\hat{Z}}$, that operates directly on node embeddings learned from signed networks.
Instead of relying on 
predefined node states, the proposed metric interprets embeddings as \textit{continuous, multi-dimensional latent states} that capture relative structural alignment induced by positive and negative relations. 
To ensure invariance to embedding dimensionality and scale while preserving sensitivity to antagonistic structure, \ours\ aligns embedding dimensions with \textit{community-level separation} and computes polarization via an \textit{effective-resistance–based formulation}.

On the \textbf{mitigation side}, \ours\ adopts a structure-aware strategy that avoids directly rewiring opposing communities or aggressively minimizing polarization scores.
Instead, \ours\ identifies a set of intermediary nodes, referred to as a \textit{gray zone}, that are structurally balanced and positioned between polarized groups.
By selectively inserting edges through these nodes, \ours\ mitigates polarization in a localized and indirect manner, enabling controlled trade-offs between polarization reduction and downstream task performance.

\vspace{1mm}
\noindent\textbf{Contributions.}
Our contributions are as follows:
\begin{itemize}[leftmargin=*]
    \item \textbf{New Perspective on Signed Network Embedding.}
    We highlight an underexplored perspective that structural polarization can intensify as a representation-level phenomenon when SNE are applied to temporal signed networks. This shifts attention from model accuracy alone to the long-term ecosystem-level consequences of embedding reuse.
    
    \item \textbf{Unified Polarization Management Framework.}
    We propose \ours, a unified framework that jointly measures and mitigates polarization directly in the signed network embedding space. 



    \item \textbf{Comprehensive Validation.}
    We conduct extensive experiments on real-world signed networks to demonstrate the effectiveness, robustness and controllability of the proposed framework.
\end{itemize}

\section{Preliminaries} \label{sec:prelim} In this section, we review an effective resistance-based formulation of polarization originally developed for \textit{unsigned} networks.
This formulation serves as the starting point of our analysis and provides the foundation upon which our embedding-aware extension is built.

\vspace{1mm}
\noindent\textbf{Effective Resistance in Networks.}
Let $\mathcal{G} = (\mathcal{V}, \mathcal{E})$ denote an undirected and unsigned network, where $\mathcal{V}$ is the set of nodes and $\mathcal{E} \subseteq \mathcal{V} \times \mathcal{V}$ is the set of edges.
For two nodes $u, v \in \mathcal{V}$, \textit{effective resistance} \cite{klein_randic93_resistance} is a graph-theoretic distance that quantifies how difficult it is for information or influence to propagate between $u$ and $v$  through $\mathcal{G}$. 
Unlike shortest-path distance, it accounts for \textit{all} possible paths between two nodes, assigning lower resistance to pairs connected through multiple redundant paths.

Formally, let $\mathbf{L} = \mathbf{D} - \mathbf{A}$ be the (combinatorial) graph Laplacian of $\mathcal{G}$, where $\mathbf{A}$ is the adjacency matrix and $\mathbf{D}$ is the diagonal degree matrix, and let $\mathbf{L}^{\dagger}$ denote its Moore-Penrose pseudoinverse~\cite{klein_randic93_resistance}.\footnote{The pseudoinverse is required since $\mathbf{L}$ is singular, and it enables effective resistance to be defined through a quadratic form over the network structure.}
The effective resistance $R(u,v)$ between $u, v \in \mathcal{V}$ is defined as:

\vspace{-0.2cm}
\small
\begin{equation}
    R(u,v) = (\mathbf{e}_u - \mathbf{e}_v)^{\top} \mathbf{L}^{\dagger} (\mathbf{e}_u - \mathbf{e}_v),
\end{equation}
\normalsize
where $\mathbf{e}_u$ and $\mathbf{e}_v$ are the indicator vectors of $u$ and $v$, respectively. 
Intuitively, $R(u,v)$ is large when $u$ and $v$ are weakly connected, and small when they are connected by many redundant paths.

\noindent\textbf{Polarization via Effective Resistance.}
Building on effective resistance, prior work \cite{hohmann_sciadv23_1}
 defines polarization by incorporating \textit{node-level opinions}. 
Each node $v \in \mathcal{V}$ is associated with a \textit{scalar opinion} $o_v \in \{-1,0,+1\}$, indicating opposition, neutrality, or support toward a given issue, which is assumed to be given exogenously.
For example, in \cite{hohmann_sciadv23_1}, $o_v$ is estimated from observed data: 
in social media networks, it is derived from the political leaning of the information sources shared by user $v$, resulting in a continuous value in $[-1, +1]$.
For expositional clarity, such continuous opinions are often discretized to preserve polarity while simplifying the analysis.
Here, $\mathbf{o} \in \mathbb{R}^{|\mathcal{V}|}$ represents the opinion vector over nodes, and $\mathbf{o}^+$ and $\mathbf{o}^-$ denote its positive and negative parts, respectively. 

Given a network $\mathcal{G}$ and the opinion vector $\mathbf{o}$, the polarization $\delta_{\mathcal{G}, \mathbf{o}}$ of $\mathcal{G}$ is computed as follows \cite{hohmann_sciadv23_1}:

\vspace{-0.2cm}
\small
\begin{equation}
\delta_{\mathcal{G}, \mathbf{o}}
= \sqrt{(\mathbf{o}^+ - \mathbf{o}^-)^{\top}
\mathbf{L}^{\dagger}
(\mathbf{o}^+ - \mathbf{o}^-)}.
\end{equation}
\normalsize
This formulation can also be interpreted as a generalized Euclidean distance induced by the effective resistance of the network \cite{coscia20_generalized_euclidean}.
Intuitively, this quantity measures how much structural resistance the network imposes when nodes with opposing opinions are placed at different positions, yielding larger values when such nodes are weakly connected and difficult to bridge through the network.

\begin{figure*}[t]
    \centering
    \includegraphics[width=0.95\textwidth]{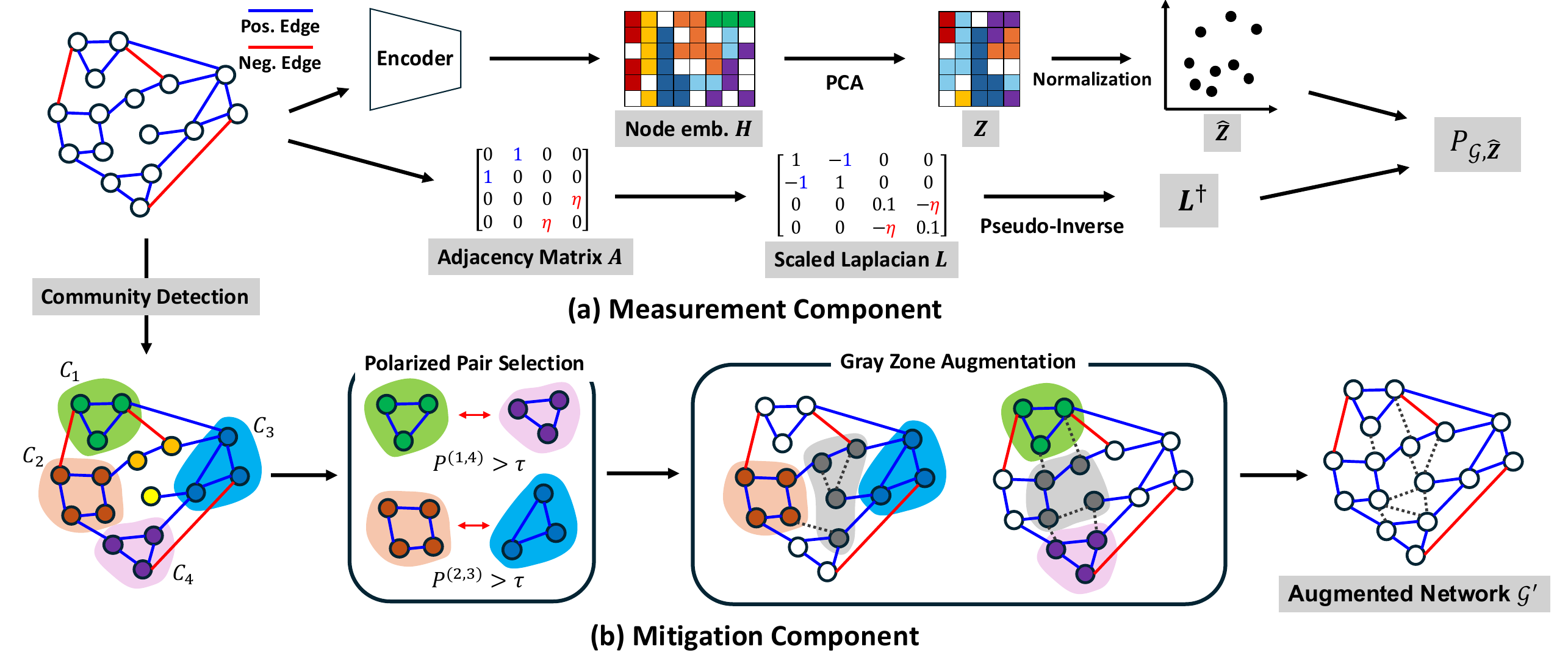}
    \vspace{-0.4cm}
    \caption{Overview of the proposed \ours\ framework.
    The framework consists of (a) an embedding-aware polarization measurement component that quantifies polarization from signed network embeddings, and (b) a gray-zone-based mitigation component, which reduces polarization via indirect structural augmentation while preserving the original network structure.}
    \label{fig:EMP_overview}
    \vspace{-0.3cm}
\end{figure*}

\vspace{1mm}
\noindent\textbf{Multi-dimensional Extensions.}
In many real-world settings, polarization does not arise along a single binary dimension but instead involves multiple competing positions \cite{delvalle2018_echo,delvalle2022_political_interaction}.
Motivated by this observation, the effective-resistance-based formulation has been extended to multi-dimensional settings \cite{hohmann_sciadv23_1}. 
A representative example is multi-party political systems, where more than two competing options coexist.
Let $\{A_1, A_2, \ldots, A_k\}$ denote a set of $k$ political options or categories.
Under this setting, prior work \cite{hohmann_sciadv23_1} associates each option $A_i$ with an opinion vector $\mathbf{o}_{A_i} \in \mathbb{R}^{|\mathcal{V}|}$, where $\mathbf{o}_{A_i}(v)$ quantifies how strongly node $v$ supports option $A_i$.

To quantify polarization across multiple options, effective-resista- nce-based structural distances are computed for every pair of options.
For a given pair $(A_i, A_j)$, polarization is measured by the quadratic form $(\mathbf{o}_{A_i} - \mathbf{o}_{A_j})^{\top} \mathbf{L}^{\dagger} (\mathbf{o}_{A_i} - \mathbf{o}_{A_j}),$ which captures how strongly supporters of the two options $A_i$ and $A_j$ are separated by the network structure.
The overall level of polarization is then obtained by aggregating these pairwise quantities across all option pairs~\cite{hohmann_sciadv23_1}:

\vspace{-0.2cm}
\small
\begin{equation}
\delta_{\mathcal{G}, \mathbf{o}}
=
\sqrt{
\binom{k}{2}^{-1}
\sum_{i=1}^{k}
\sum_{j=i+1}^{k}
(\mathbf{o}_{A_i} - \mathbf{o}_{A_j})^{\top}
\mathbf{L}^{\dagger}
(\mathbf{o}_{A_i} - \mathbf{o}_{A_j})
}.
\label{eq:multidim_polarization}
\end{equation}
\normalsize
Intuitively, this formulation treats each pair of options as a binary polarization scenario and averages the corresponding effective-resistance-based distances.
As a result, the multidimensional measure preserves the same interpretation as the unidimensional case, quantifying how strongly the network structure separates individuals holding competing positions across multiple dimensions.



\vspace{1mm}
\noindent\textbf{Limitations in Signed and Embedding-based Settings.}
While effective-resistance-based polarization measures provide a principled way to quantify structural separation given explicit opinions, their assumptions warrant careful consideration when applied to signed and embedding-based settings.
Both the unidimensional and multidimensional formulations treat opinions as externally given node attributes and define polarization over unsigned network structures.
As a result, polarization is characterized with respect to predefined opinion vectors, without explicitly accounting for signed relations or representations learned from the network itself.

Moreover, the multidimensional extension assumes that each individual supports exactly one option and that polarization across multiple dimensions can be captured by aggregating pairwise separations between options.
Such assumptions contrast with embedding-based analyses, where node representations are continuous, latent, and not naturally associated with mutually exclusive categories.
These observations suggest that directly applying existing effective-resistance-based formulations to SNE may be nontrivial, motivating a closer examination of how polarization should be defined in representation-learning settings.


\section{The Proposed Framework: \ours} \label{sec:method}

In this section, we present \textbf{\ours},
an embedding-aware framework for measuring and mitigating polarization in signed networks.
We first provide an overview of \ours\ (\textsection\ref{sec:overview}), and then introduce an embedding-aware polarization measurement method (\textsection\ref{sec:measure}) followed by a structure-aware mitigation strategy (\textsection\ref{sec:miti}).

\begin{table}[t]
\footnotesize
\centering
\caption{Key notations used in this paper}
\label{tab:notation}
\vspace{-0.25cm}
\renewcommand{\arraystretch}{1.3}
\begin{tabular}{c|p{0.76\columnwidth}}
\hline
\textbf{Notation} & \textbf{Description} \\
\hline
$\mathcal{G}$, $\mathcal{G}'$
& Original and augmented signed networks \\
$\mathcal{V}, \mathcal{E}^+, \mathcal{E}^-$
& Node set, positive edges, and negative edges\\
$\mathcal{E}^{\mathrm{gray}}$
& Gray-zone edges added for polarization mitigation \\
\hline
$\mathbf{H},\ \hat{\mathbf{Z}}$ 
& Raw and normalized node embedding matrices \\
$P_{\mathcal{G},\hat{\mathbf{Z}}}$ 
& Embedding-aware polarization score of $\mathcal{G}$ \\
$P^{(i,j)}$ 
& Polarization score between communities $C_i$ and $C_j$ \\
\hline
$\eta$ 
& Scaling factor for negative edges in the Laplacian \\
$\gamma$ 
& Mitigation strength parameter \\
$\tau,\ d_{\max}$ 
& Pair-selection threshold and maximum selections per community \\
\hline
\end{tabular}
\vspace{-0.4cm}
\end{table}

\subsection{Overview}\label{sec:overview}
We consider a signed network $\mathcal{G} = (\mathcal{V}, \mathcal{E}^+, \mathcal{E}^-)$, where $\mathcal{V}$ denotes the set of nodes, and $\mathcal{E}^+$ and $\mathcal{E}^-$ represent positive and negative edges, respectively.
Given $\mathcal{G}$, a signed network embedding (SNE) model learns a node representation matrix $\mathbf{H} \in \mathbb{R}^{|\mathcal{V}| \times d}$, where each row $\mathbf{h}_v$ encodes the latent structural role of node $v$ induced by signed relations.
Our goal is to measure and mitigate structural polarization directly in this embedding space, without relying on externally specified opinions or predefined node states.


Figure~\ref{fig:EMP_overview} illustrates the overall pipeline of \ours, and Table~\ref{tab:notation} summarizes a list of notations used in this paper.
The framework consists of two components: an \textbf{embedding-aware measurement} component (Figure~\ref{fig:EMP_overview}-(a)) and a \textbf{structure-aware mitigation} component (Figure~\ref{fig:EMP_overview}-(b)).
Together, these components are tightly coupled, forming a unified framework for polarization management in embedding-based signed network analysis.

On the \textbf{measurement component}, \ours\ takes as input a signed network $\mathcal{G}$ and its corresponding node embeddings $\mathbf{H}$ learned by an SNE model.
The embeddings are first transformed into a normalized representation $\hat{\mathbf{Z}}$ to ensure comparability across dimensions.
Each embedding dimension is then interpreted as a latent axis of structural alignment induced by signed relations.
To quantify polarization along each axis, \ours\ combines $\hat{\mathbf{Z}}$ with a \textit{scaled signed Laplacian} of $\mathcal{G}$, which encodes how positive and negative edges constrain information flow over the network.
Based on this formulation, an \textit{effective resistance (ER)-based energy} is computed for each dimension, and these quantities are aggregated to produce a single network-level polarization score $P_{\mathcal{G}, \hat{\mathbf{Z}}}$.

On the \textbf{mitigation component}, \ours\ takes as input the signed network $\mathcal{G}$, node embeddings $\mathbf{H}$, and normalized embeddings $\hat{\mathbf{Z}}$.
\ours\ identifies communities in $\mathcal{G}$ using $\mathbf{H}$, and computes a polarization score $P^{(i,j)}$ for each community pair $(C_i, C_j)$ based on their separation in the embedding space.
A subset of \textit{highly polarized pairs} is then selected as mitigation targets.
For each selected pair, \ours\ constructs a \textit{gray zone} consisting of intermediary nodes that are structurally balanced and proximate to both communities.
Finally, \ours\ yields an augmented signed network $\mathcal{G}'$ by selectively inserting a small number of edges through these gray-zone nodes.
Polarization can then be re-evaluated on $\mathcal{G}'$ for analysis, although the mitigation procedure itself does not require repeated scoring.




\subsection{Polarization Measurement}\label{sec:measure}
The goal of this component is to quantify polarization in signed networks from an embedding-based perspective. 
As discussed in Section~\ref{sec:prelim}, existing ER–based formulations are primarily developed for unsigned networks and assume discrete opinion states, which are not directly compatible with continuous representations learned from signed networks.
To bridge this gap, \ours\ reformulates ER-based polarization to operate on embedding geometry rather than predefined opinion states.
This component proceeds in two stages, each of which is described in detail below.

\vspace{1mm}
\noindent\textbf{Stage 1: Latent Structural Alignment.}
This stage constructs a latent structural space in which polarization can be consistently measured across embedding dimensions. 
Although node embeddings learned by SNE models capture structural patterns in signed networks, they are optimized for representation objectives rather than quantitative measurement, and thus do not inherently guarantee comparability or scale invariance.

Accordingly, \ours\ aligns and normalizes the learned embeddings to construct a coherent measurement space for dimension-wise polarization analysis.
To this end, the aligned latent space is designed to satisfy the following requirements:
\begin{itemize}[leftmargin=*]
    \item \textbf{Dimension-wise comparability:} Embedding dimensions are directly comparable for aggregating polarization contributions.
    \item \textbf{Salient structural separation:} Redundant directions are suppressed to emphasize dominant antagonistic patterns.
    \item \textbf{Scale invariance:} The representation is invariant to embedding scale, ensuring polarization reflects intrinsic network structure.
\end{itemize}

We begin with node embeddings $\mathbf{H} = f_{\theta}(\mathcal{G}) \in \mathbb{R}^{|\mathcal{V}| \times d}$ learned from an SNE model; 
in this work, we employ \textit{signed graph convolutional networks} (SGCN) \cite{derr_sgcn_icdm18} as a representative model.
In practice, fixing the embedding dimensionality \textit{a priori} risks collapsing meaningful antagonistic patterns or introducing redundant dimensions, as polarization in signed networks may emerge along multiple structural separation modes.
To adaptively set the dimensionality, \ours\ estimates the \textbf{target dimensionality $k$} using a set of \textit{preliminary communities} detected directly from the signed topology. 
Specifically, we apply a \textit{signed Louvain} method \cite{gomez2009_signed_community} to obtain a coarse-grained partition $\mathcal{C}^{\mathrm{pre}}$ of $\mathcal{G}$, and set the target dimensionality as $k = \lvert \mathcal{C}^{\mathrm{pre}} \rvert$.
These preliminary communities are used \textit{only} to approximate the number of dominant structural separation modes; 
the community assignments themselves are not used.

Given the target dimensionality $k$, \ours\ aligns the learned embeddings along dominant structural separation axes.
To this end, we construct an \textbf{aligned latent representation} $\mathbf{Z}$ by applying \textit{principal component analysis (PCA)} \cite{jolliffe1990_pca} to $\mathbf{H}$:

\vspace{-0.2cm}
\small
\begin{equation}
\mathbf{Z}= \mathbf{H}\mathbf{W}, \quad \mathbf{Z} \in \mathbb{R}^{|\mathcal{V}| \times k},
\end{equation}
\normalsize
where $\mathbf{W} \in \mathbb{R}^{d \times k}$ consists of the top-$k$ eigenvectors of the covariance matrix of $\mathbf{H}$.
Here, PCA aligns embedding dimensions with the dominant structural separation modes in the signed network, suppressing redundant directions and yielding a representation suitable for dimension-wise polarization analysis.

While PCA aligns embedding dimensions, the absolute scale of $\mathbf{Z}$ may still vary across models and datasets, which can arbitrarily influence polarization measurements.
To ensure scale invariance, \ours\ applies global \textit{root mean square} (RMS) normalization to $\mathbf{Z}$ and obtains the \textbf{normalized representation} $\hat{\mathbf{Z}}$ as follows:

\vspace{-0.2cm}
\small
\begin{equation}
\hat{\mathbf{Z}}
=
\frac{\mathbf{Z}}
{\sqrt{\frac{1}{|\mathcal{V}|k}
\sum_{u=1}^{|\mathcal{V}|} \sum_{\ell=1}^{k} z_{u\ell}^2}}, \quad \hat{\mathbf{Z}} \in \mathbb{R}^{|\mathcal{V}| \times k}.
\end{equation}
\normalsize
The normalized matrix $\hat{\mathbf{Z}}$ defines the aligned latent structural space for polarization measurement, where each dimension corresponds to a latent axis of structural separation induced by signed relations.

\vspace{1mm}
\noindent\textbf{Stage 2: Polarization Computation in Embedding Space.}
This stage defines an embedding-aware polarization measure $P_{\mathcal{G},\hat{\mathbf{Z}}}$ by combining the normalized embeddings with signed structural constraints in $\mathcal{G}$.
The proposed measure builds on effective resistance but departs fundamentally from prior ER-based formulations by operating on an aligned, continuous embedding geometry rather than on discrete opinion assignments or external labels.

Effective resistance is defined through the Moore-Penrose pseudo-inverse of a positive semi-definite (PSD) Laplacian.
To incorporate signed relations while preserving this property, we construct a scaled Laplacian in which the contribution of negative edges is modulated by a factor $\eta\in(0,1)$.
To this end, we first define a scaled adjacency matrix $\mathbf{A}\in\mathbb{R}^{|\mathcal{V}|\times|\mathcal{V}|}$ as follows:

\vspace{-0.2cm}
\footnotesize
\begin{equation}
A_{uv}=
\begin{cases}
1, & \text{if } (u,v)\in\mathcal{E}^+,\\
\eta, & \text{if } (u,v)\in\mathcal{E}^-,\\
0, & \text{otherwise.}
\end{cases}
\end{equation}
\normalsize
That is, positive edges are assigned weight $1$, while negative edges are down-weighted by a factor $\eta$ that controls the strength of antagonistic influence encoded by negative edges.
Intuitively, smaller values of $\eta$ amplify structural separation by increasing effective resistance, whereas larger values attenuate this influence, making the signed network geometry closer to that of an unsigned network.

Based on this scaled adjacency, we define the weighted degree of $u$ as $d_u=\sum_{v}A_{uv}$ and let $\mathbf{D}=\mathrm{diag}(d_1,\ldots,d_{|\mathcal{V}|})$.
The corresponding Laplacian $\mathbf{L}$ of $\mathcal{G}$ is then defined as $\mathbf{L}=\mathbf{D}-\mathbf{A}$.
This construction preserves positive semi-definiteness, while allowing negative relations to influence the induced effective-resistance geometry through $\eta$.
Equivalently, the $(u,v)$-th entry of $\mathbf{L}$ is given by:

\vspace{-0.2cm}
\footnotesize
\begin{equation}
L_{uv} =
\begin{cases}
d_u, & \text{if } u = v, \\
-1, & \text{if } (u,v) \in \mathcal{E}^+, \\
-\eta, & \text{if } (u,v) \in \mathcal{E}^-, \\
0, & \text{otherwise.}
\end{cases}
\end{equation}
\normalsize
Recall that $\eta$ controls the extent to which negative relations contribute to the effective-resistance geometry induced by $\mathbf{L}$.
We examine its impact empirically in Section~\ref{sec:eval} by analyzing how changes in $\eta$ affect measured polarization.

We now quantify polarization directly in the embedding space.
We interpret the normalized embeddings $\hat{\mathbf{Z}}=[\hat{\mathbf{z}}_1,\ldots,\hat{\mathbf{z}}_{k}] \in \mathbb{R}^{|\mathcal{V}|\times k}$ obtained in Stage 1 by treating each column $\hat{\mathbf{z}}_\ell\in\mathbb{R}^{|\mathcal{V}|}$ as an independent latent axis of structural separation.
This column-wise interpretation enables polarization to be measured across multiple modes without imposing discrete opinion categories, treating each dimension as a continuous latent coordinate rather than a categorical pole. 
In other words, explicit pairwise opinion contrasts in Eq.~(\ref{eq:multidim_polarization}) are replaced by an axis-wise, structure-aware energy formulation.

Specifically, for each dimension $\ell$, we define the effective-resistance energy induced by the aligned embedding axis $\hat{\mathbf{z}}_\ell$ on $\mathcal{G}$ as follows:

\vspace{-0.2cm}
\footnotesize
\begin{equation}
\mathcal{E}_\ell(\mathcal{G},\hat{\mathbf{Z}})
=
\hat{\mathbf{z}}_\ell^{\top}\mathbf{L}^{\dagger}\hat{\mathbf{z}}_\ell,
\end{equation}
\normalsize
where $\mathbf{L}^{\dagger}$ denotes the Moore--Penrose pseudo-inverse of the Laplacian of $\mathcal{G}$.
The quantity $\mathcal{E}_\ell(\mathcal{G},\hat{\mathbf{Z}})$ represents the structural energy associated with variation along the $\ell$-th latent axis, and serves as the fundamental unit of polarization prior to aggregation.

We then aggregate the dimension-wise energies to obtain a single \textbf{network-level polarization score} as follows:

\vspace{-0.2cm}
\small
\begin{equation}
P_{\mathcal{G},\hat{\mathbf{Z}}}
=
\sqrt{
\frac{1}{k}
\sum_{\ell=1}^{k}
\hat{\mathbf{z}}_\ell^{\top}
\mathbf{L}^{\dagger}
\hat{\mathbf{z}}_\ell
}.
\end{equation}
\normalsize
That is, we average across dimensions to avoid trivial scaling with the embedding dimensionality, and apply a square root to keep $P_{\mathcal{G},\hat{\mathbf{Z}}}$ on the same scale as effective-resistance distances.

\subsection{Polarization Mitigation}\label{sec:miti}
The goal of this component is to mitigate polarization in signed networks through targeted structural augmentation.
Rather than applying global rewiring or directly bridging opposing communities, \ours\ leverages the measured polarization to guide \textit{where} and \textit{how} structural modification should be applied.
This component proceeds in three stages, each of which is described in detail below.



\vspace{1mm}
\noindent\textbf{Stage 1: Polarized Community Pair Identification.}
This stage identifies pairs of communities for which mitigation interventions are expected to be most effective.
Empirical studies on signed networks suggest that polarization is often dominated by pronounced structural separation between a small number of community pairs, rather than being uniformly distributed across the entire network \cite{bonchi_cikm19_1,huang_pole_wsdm22}.
Motivated by this observation, \ours\ localizes such community-level antagonism and treats highly polarized community pairs as primary mitigation targets.

We begin by partitioning the node set $\mathcal{V}$ into $k$ communities using $k$-means clustering~\cite{macqueen1967_kmeans} on the raw node embedding matrix $\mathbf{H}$ learned by the SNE model. 
We use $\mathbf{H}$ for this step because it directly preserves structural positions,
whereas the aligned embedding $\hat{\mathbf{Z}}$ is tailored for polarization measurement
and emphasizes dominant polarization directions rather than fine-grained community structure.
The value of $k$ is set to match that used in Section~\ref{sec:measure}.
While earlier communities are used only to determine $k$, the communities constructed here serve directly as the units for mitigation.
This yields a community partition
$\mathcal{C} = \{C_1, \ldots, C_k\}$ with $C_i \subset \mathcal{V}$.

For each unordered community pair $(C_i, C_j)$, we assess their mutual antagonism by constructing a community-restricted view of the network and embedding space.
Let $\hat{\mathbf{Z}} \in \mathbb{R}^{|\mathcal{V}|\times k}$ denote the aligned and normalized embedding matrix introduced in Section~\ref{sec:measure}.
For a given pair $(C_i, C_j)$, we restrict $\hat{\mathbf{Z}}$ to the node subset $C_i \cup C_j$, yielding the corresponding submatrix  $\hat{\mathbf{Z}}_{i,j}$, and consider the induced subgraph $\mathcal{G}_{i,j} = \mathcal{G}[C_i \cup C_j]$.
Before computing polarization, we apply mean-centering to $\hat{\mathbf{Z}}_{i,j}$ to stabilize ER computation.

Using the restricted embeddings $\hat{\mathbf{Z}}_{i,j}$, we then apply the embedding-aware polarization measure to $\mathcal{G}_{i,j}$ .
The resulting \textbf{pairwise polarization score} for the community pair $(C_i, C_j)$ is defined as follows:

\vspace{-0.2cm}
\footnotesize
\begin{equation}
P^{(i,j)} \;\triangleq\;
P_{\mathcal{G}_{i,j},\, \hat{\mathbf{Z}}_{i,j}}.
\end{equation}
\normalsize

To select community pairs for mitigation from the set of pairwise polarization scores
$\{P^{(i,j)}\}$, we use the following hyperparameters:
\begin{itemize}[leftmargin=*]
    \item \textbf{Polarization threshold} $\boldsymbol{\tau}$: We apply min-max normalization to $\{P^{(i,j)}\}$ and select community pairs whose normalized scores exceed $\tau$.
    This ensures that structural intervention is applied only to sufficiently polarized pairs.
    \item \textbf{Maximum selection count} $\boldsymbol{d_{\max}}$: To prevent mitigation from being dominated by the same communities, we cap the number of times each community can appear in selected pairs at $d_{\max}$.
\end{itemize}
Accordingly, the resulting set of selected community pairs is denoted by $\mathcal{P}$.
We empirically analyze the effect of these hyperparameters in Section~\ref{sec:eval}, demonstrating how they control the trade-off between mitigation effectiveness and structural preservation.

\vspace{1mm}
\noindent\textbf{Stage 2: Gray Zone Identification.}
This stage identifies a set of intermediary nodes that enable indirect structural interaction between polarized community pairs, without directly connecting the communities themselves.
The key idea is to mitigate polarization by introducing \textit{buffering structures} that reduce structural separation while preserving the original network topology.

Given a selected community pair $(C_i, C_j) \in \mathcal{P}$, we consider nodes $v \in \mathcal{V} \setminus (C_i \cup C_j)$, \ie, nodes outside the two communities, as candidates for gray-zone nodes.
An effective gray node should satisfy two complementary criteria: 
(i) it should be \textit{proximal} to the two communities in the aligned embedding space, reflecting geometric closeness, and
(ii) it should be \textit{balanced} with respect to the two communities, reflecting symmetric relations.

To formalize these criteria, we quantify the proximity of a candidate node $v$ is to each community $C_i$ as follows:

\vspace{-0.2cm}
\footnotesize
\begin{equation}
d_i(v) = \lVert \mathbf{z}_v - \boldsymbol{\mu}_i \rVert,
\qquad
\boldsymbol{\mu}_i = \frac{1}{|C_i|} \sum_{u \in C_i} \mathbf{z}_u,
\end{equation}
\normalsize
where $\boldsymbol{\mu}_i$ denotes the community centroid of $C_i$.
The proximity $d_j(v)$ to community $C_j$ is defined analogously.
Using these distances, we define a \textbf{gray distance} $s(v \mid C_i, C_j)$ for node $v$, measuring its suitability as an intermediary between the community pair $(C_i, C_j)$:

\vspace{-0.2cm}
\footnotesize
\begin{equation}
s(v \mid C_i, C_j)
=
\big| d_i(v) - d_j(v) \big|
+
\max \{ d_i(v), d_j(v) \} .
\end{equation}
\normalsize
The first term penalizes imbalance between the two communities and reflects the balanced criterion, whereas the second term captures absolute distance and reflects proximity to both communities.
Smaller values of $s(v \mid C_i, C_j)$ therefore correspond to nodes that are both well-balanced and structurally close to the polarized pair.

Based on this distance, we select a fixed-size \textbf{gray-zone node set} $\mathcal{V}^{\mathrm{gray}}_{(i,j)}$ for $(C_i, C_j)$ as follows:

\vspace{-0.2cm}
\footnotesize
\begin{equation}
\mathcal{V}^{\mathrm{gray}}_{(i,j)}
= \operatorname{Bottom\text{-}}m \big( s(v \mid C_i, C_j) \big), \\ \qquad
m
= \left\lceil \frac{1}{|\mathcal{C}|} \sum_{r=1}^{|\mathcal{C}|} |C_r| \right\rceil ,
\end{equation}
\normalsize
where $\operatorname{Bottom\text{-}}m(\cdot)$ returns the $m$ candidate nodes with the \textit{smallest} gray distances, and $|C_r|$ denotes the number of nodes in community $C_r$.
That is, we set $m$ to the average community size to align the scale of gray-zone intervention with typical community-level structures, avoiding both negligible and overly aggressive augmentation.

\vspace{1mm}
\noindent\textbf{Stage 3: Gray-Zone–Aware Augmentation.}
This stage performs a controlled structural augmentation based on polarized community pairs and their associated gray-zone node sets. 
To this end, \ours\ inserts a limited number of \textit{positive} edges involving gray-zone nodes.
The intuition behind this design is to mitigate polarization while minimizing the risk of introducing additional antagonism through negative-edge augmentation.
Although structural augmentation can modify the network structure, adding positive edges through gray-zone nodes introduces indirect interaction pathways without explicitly creating new antagonistic ties.
We leave the exploration of mitigation strategies involving negative-edge augmentation as an interesting direction for future research.

To balance mitigation and structural distortion, we first regulate the augmentation scale using dataset-level community connectivity statistics.
Specifically, we summarize the typical volume of positive interactions \textit{within} and \textit{between} communities in $\mathcal{G}$:

\vspace{-0.2cm}
\footnotesize
\begin{equation}
\bar{e}^{+}_{\mathrm{intra}}
=
\frac{1}{|\mathcal{C}|}
\sum_{r=1}^{|\mathcal{C}|}
\big| \mathcal{E}^{+}(C_r, C_r) \big|, \quad
\bar{e}^{+}_{\mathrm{inter}}
=
\frac{2}{|\mathcal{C}|(|\mathcal{C}|-1)}
\sum_{1 \le r < s \le |\mathcal{C}|}
\big| \mathcal{E}^{+}(C_r, C_s) \big|,
\end{equation}
\normalsize
where $\mathcal{E}^{+}(C_r, C_s)$ denotes the set of positive edges connecting nodes in
communities $C_r$ and $C_s$.
Accordingly, $\bar{e}^{+}_{\mathrm{intra}}$ reflects typical \textit{intra-community} positive connectivity, whereas $\bar{e}^{+}_{\mathrm{inter}}$ captures average \textit{inter-community} positive connectivity.

Using these statistics, we allocate an \textbf{augmentation budget} for each pair $(C_i, C_j) \in \mathcal{P}$.
Specifically, the budget is distributed across three types of positive connections, parameterized by $(b_{gg}, b_{gi}, b_{gj})$:
(i) among gray-zone nodes, (ii) between the gray zone and community $C_i$, and (iii) between the gray zone and community $C_j$.

\vspace{-0.2cm}
\footnotesize
\begin{equation}
b_{gg}=
\big\lfloor \gamma \, \bar{e}^{+}_{\mathrm{intra}} \big\rfloor, \quad
b_{gi} = b_{gj}=
\left\lfloor \frac{\gamma \, \bar{e}^{+}_{\mathrm{inter}}}{2} \right\rfloor,
\end{equation}
\normalsize
where $b_{gg}$ specifies the number of positive edges inserted among gray-zone nodes,
while $b_{gi}$ and $b_{gj}$ specify the numbers of edges connecting the gray zone to communities $C_i$ and $C_j$, respectively.
The scaling parameter $\gamma > 0$ acts as a global intervention knob, controlling the \textbf{trade-off between polarization mitigation and structural distortion}.
Smaller values of $\gamma$ yield \textit{conservative} interventions that largely preserve the original network structure, whereas larger values enable more \textit{aggressive} augmentation to alleviate structural separation.
We examine this trade-off in Section~\ref{sec:eval}.

Given the allocated budgets $(b_{gg}, b_{gi}, b_{gj})$, we construct the structural augmentation by adding new positive edges between selected node pairs.
Specifically, we randomly sample $b_{gg}$ unconnected node pairs among gray-zone nodes, and $b_{gi}$ and $b_{gj}$ unconnected pairs between the gray zone and community $C_i$ and community $C_j$, respectively.
To prevent duplicating existing structure, we exclude candidate edges that already exist in the original graph.
As a result, the gray-zone–aware augmentation with a polarized community pair $(C_i, C_j)$
yields an intermediate augmented graph defined as:

\vspace{-0.2cm}
\footnotesize
\begin{equation}
\mathcal{G}'_{(i,j)}=\big( \mathcal{V},\; \mathcal{E} \cup \mathcal{E}^{\mathrm{gray}}_{(i,j)} \big), \qquad
\mathcal{E}^{\mathrm{gray}}_{(i,j)} = \mathcal{E}_{gg} \cup \mathcal{E}_{gi} \cup \mathcal{E}_{gj},
\end{equation}
\normalsize
where $\mathcal{E}_{gg}$, $\mathcal{E}_{gi}$, and $\mathcal{E}_{gj}$ denote the resulting sets of sampled edges, with $|\mathcal{E}_{gg}| = b_{gg}$, $|\mathcal{E}_{gi}| = b_{gi}$, and $|\mathcal{E}_{gj}| = b_{gj}$, respectively.

Aggregating augmentations across all polarized community pairs in $\mathcal{P}$, we obtain the final augmented network $\mathcal{G}'$ as follows:

\vspace{-0.2cm}
\footnotesize
\begin{equation}
\mathcal{G}'
=
\Big(
\mathcal{V},\;
\mathcal{E}
\;\cup\;
\mathcal{E}^{\mathrm{gray}}
\Big), \qquad 
\mathcal{E}^{\mathrm{gray}}
=
\bigcup_{(C_i,C_j) \in \mathcal{P}}
\mathcal{E}^{\mathrm{gray}}_{(i,j)}.
\end{equation}
\normalsize

By leveraging gray-zone nodes for indirect structural augmentation, \ours\ reduces structural separation between polarized communities while preserving the original network connectivity.

\section{Evaluation} \label{sec:eval} 
\begin{table}[!t]
\small
\centering
\caption{Dataset statistics}
\vspace{-0.25cm}
\label{tab:dataset_stats}
\resizebox{0.8\columnwidth}{!}{ 
\begin{tabular}{lrrrr}
\toprule
\textbf{Dataset} & \textbf{\#Nodes} & \textbf{\#Edges} & \textbf{\#Positive} & \textbf{\#Negative} \\
\midrule
\textbf{BTC-Alpha}   & 3{,}783   & 14{,}124   & 12{,}947   & 1{,}177 \\
\textbf{BTC-OTC}     & 5{,}881   & 21{,}492   & 18{,}574   & 2{,}918 \\
\textbf{Wiki-Elec}   & 7{,}115   & 100{,}762  & 78{,}639   & 22{,}123 \\
\textbf{Wiki-RfA}    & 11{,}256  & 170{,}757  & 133{,}271  & 37{,}486 \\
\textbf{Slashdot}    & 82{,}140  & 500{,}481  & 381{,}648  & 118{,}833 \\
\textbf{Epinions}    & 131{,}580 & 711{,}210  & 592{,}013  & 119{,}197 \\
\bottomrule
\end{tabular}
\vspace{-0.25cm}
}
\end{table}

\begin{table*}[t]
\centering
\caption{Effect of gray-zone-based mitigation on predictive performance and polarization (EQ2). Gray-zone augmentation yields substantial polarization reduction across all datasets, while maintaining predictive performance within a controlled range.}
\label{tab:eq2_mitigation}
\vspace{-0.2cm}
\footnotesize
\setlength{\tabcolsep}{9pt}
\resizebox{0.92\textwidth}{!}{ 
\begin{tabular}{l|ccc|ccc|ccc}
\hline
\multirow{2}{*}{\textbf{Dataset}}
& \multicolumn{3}{c|}{\textbf{Accuracy}}
& \multicolumn{3}{c|}{\textbf{Macro-F1}}
& \multicolumn{3}{c}{\textbf{Polarization $\boldsymbol{P_{\mathcal{G},\hat{\mathbf{Z}}}}$}} \\
\cline{2-10}
& \textbf{Baseline} & \textbf{\ours} & $\boldsymbol{\Delta \%}$
& \textbf{Baseline} &  \textbf{\ours} & $\boldsymbol{\Delta \%}$
& \textbf{Baseline} &  \textbf{\ours} & $\boldsymbol{\Delta \%}$ \\
\hline
\textbf{BTC-Alpha}
& 0.745 & 0.751 & +0.84
& 0.697 & 0.691 & -0.90
& 72.40 & 49.88 & -31.11 \\
\textbf{BTC-OTC}
& 0.795 & 0.742 & -6.76
& 0.714 & 0.656 & -8.17
& 93.52 & 65.04 & -30.46 \\
\textbf{Wiki-Elec}
& 0.771 & 0.726 & -5.86
& 0.673 & 0.611 & -9.25
& 83.98 & 60.87 & -27.52 \\
\textbf{Wiki-RfA}
& 0.869 & 0.849 & -2.21
& 0.740 & 0.735 & -0.75
& 99.01 & 62.41 & -36.96 \\
\textbf{Slashdot}
& 0.756 & 0.755 & -0.06
& 0.713 & 0.711 & -0.32
& 474.44 & 412.62 & -13.03 \\
\textbf{Epinions}
& 0.844 & 0.840 & -0.58
& 0.788 & 0.787 & -0.09
& 663.23 & 515.49 & -22.28 \\
\hline
\end{tabular}
}
\vspace{-0.1cm}
\end{table*}

To evaluate effectiveness of \ours, we designed our experiments, aiming at answering the following key evaluation questions (EQs): 

\begin{itemize}[leftmargin=*]
    \item \textbf{(EQ1)} 
    Does the proposed polarization measure reliably reflect controlled changes in antagonistic structure in signed networks?
    
    \item \textbf{(EQ2)} 
    Can gray-zone-based mitigation substantially reduce polarization while maintaining predictive performance?
    
    \item \textbf{(EQ3)} 
    How does mitigation strength $\gamma$ affect the trade-off between polarization reduction and predictive performance?
    
    \item \textbf{(EQ4)} 
    How sensitive is the mitigation outcome to the pair-selection hyperparameters $\tau$ and $d_{\max}$?

    \item \textbf{(EQ5)} 
    Is \ours\ computationally efficient in practice?
\end{itemize}

\subsection{Experimental Setup}
\label{sec:exp_setup}

\textbf{Datasets.}
We conduct experiments on six real-world signed network datasets: BTC-Alpha, BTC-OTC, Wiki-Elec, Wiki-RfA, Slashdot, and Epinions.
All datasets are publicly available.\footnote{https://snap.stanford.edu/data/\#signnets.}
Dataset statistics are summarized in Table~\ref{tab:dataset_stats}.
For each dataset, observed edges are split into training, validation, and test sets with a ratio of 70\%/15\%/15\%.
Following~\cite{klein_randic93_resistance}, all experiments are conducted on the largest connected component of the training graph, as ER-based measures are well-defined only for connected graphs. 
The number of communities $k$ is determined using signed Louvain on the training graph in Section~\ref{sec:measure}, yielding $k=12,11,7,6,26,$ and $27$ for BTC-Alpha, BTC-OTC, Wiki-Elec, Wiki-RfA, Slashdot, and Epinions, respectively. 
This value of $k$ is then used for $k$-means clustering to obtain community assignments in Section~\ref{sec:miti}.

\vspace{1mm}
\noindent\textbf{Evaluation Tasks and Metrics.}
We evaluate our framework along two complementary dimensions.
For \textit{polarization analysis}, we quantify structural polarization using the proposed measure $P_{\mathcal{G}, \hat{\mathbf{Z}}}$ computed from learned embeddings.
We report both absolute values and relative changes to assess the effect of structural intervention. 
For \textit{predictive utility}, we evaluate downstream performance via signed link prediction, formulated as a three-class classification task over $\{\mathcal{E}^+\text{, }\mathcal{E}^-\text{, non-edge}\}$ relations following prior SNE studies \cite{derr_sgcn_icdm18,sharma2023_dynamic_signed,kim_polardsn_cikm24}. 
Positive ($\mathcal{E}^+$) and negative ($\mathcal{E}^-$) edges are sampled from observed links, while non-edges are uniformly sampled from unconnected node pairs. 
We report Accuracy and Macro-F1 on the test set, along with relative changes after mitigation.
    

\vspace{1mm}
\noindent\textbf{Evaluation Protocols.}
Our experiments consider both polarization measurement and mitigation. 
For \textit{polarization measurement}, we compute the proposed measure $P_{\mathcal{G},\hat{\mathbf{Z}}}$ on the training graph using embeddings learned from the \textit{original} signed network $\mathcal{G}$. 
To address EQ1, we apply a controlled scaling factor $\eta$ to negative edges to systematically vary antagonistic structure, and examine whether $P_{\mathcal{G},\hat{\mathbf{Z}}}$ responds consistently to such changes. 

For \textit{polarization mitigation}, we apply the proposed intervention to the training graph $\mathcal{G}$ and re-learn embeddings from scratch on the augmented graph $\mathcal{G}'$.
To address EQ2, we compare polarization and signed link prediction performance between the original graph $\mathcal{G}$ and the augmented graph $\mathcal{G}'$. 
For EQ3 and EQ4, we further analyze how these quantities vary on $\mathcal{G}'$ as we adjust the mitigation strength $\gamma$ and the pair-selection hyperparameters $(\tau, d_{\max})$, respectively.

\vspace{1mm}
\noindent\textbf{Implementation Details.}
All reported results are averaged over 5 random seeds. 
For polarization \textit{measurement}, node embeddings are learned using SGCN~\cite{derr_sgcn_icdm18}, with model hyperparameters (\eg, hidden dimension, learning rate, and the number of layers) tuned on the validation set of each dataset; 
the selected configuration is then fixed and used consistently across all experiments.
The \textit{mitigation} hyperparameters, including $\gamma$, $\tau$, and $d_{\max}$, are also tuned on the validation set via grid search to achieve substantial polarization reduction while preserving predictive performance as measured by Macro-F1.
For EQ2, we report results obtained using these optimal mitigation configurations, while the sensitivity to $\gamma$, $\tau$, and $d_{\max}$ is further examined in EQ3 and EQ4. 
For full reproducibility, we provide complete implementation details, including search ranges, selected values, and large-scale processing, in \textbf{Appendix}.

\begin{figure}[t]
    \centering
    \captionsetup{type=figure,skip=4pt}\includegraphics[width=\columnwidth]{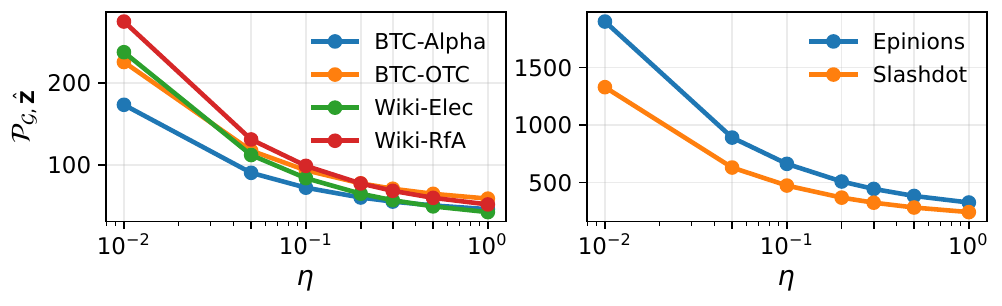}
    \caption{Effect of the negative edge scale factor $\boldsymbol{\eta}$ on polarization. Lower $\boldsymbol{\eta}$ strengthens negative edges in the ER geometry.}
    \label{fig:eq1} 
    \vspace{-0.35cm}
\end{figure}

\subsection{Results and Analysis}
\textbf{Results for EQ1.}
Figure~\ref{fig:eq1} reports the polarization score $P_{\mathcal{G}, \hat{\mathbf{Z}}}$ ($y$-axis) as a function of the negative edge scale factor $\eta$ ($x$-axis) across all datasets.
Recall that $\eta$ controls how negative edges are treated: smaller values emphasize antagonistic separation, while larger values make negative edges behave like weak positive connections

As shown in Figure~\ref{fig:eq1}, polarization consistently increases as $\eta$ decreases across all datasets, indicating that the proposed measure responds monotonically to controlled strengthening of antagonistic structure.
This confirms that our formulation realizes its intended design: the influence of negative relations can be continuously calibrated to reflect dataset-specific interpretations of antagonism.
Accordingly, we fix $\eta = 0.1$ in all subsequent experiments.

\vspace{1mm}
\noindent\textbf{Results for EQ2.}
Table~\ref{tab:eq2_mitigation} summarizes the impact of gray-zone-based mitigation on both predictive performance and polarization.
Across all datasets, \ours\ achieves substantial polarization reductions, with relative decreases ranging from 13.03\% to 36.96\%.
At the same time, degradation in downstream performance remains controlled, even on datasets exhibiting large polarization reduction.

Moderate drops in predictive accuracy are observed on some datasets (\eg, BTC-OTC, and Wiki-Elec), but this behavior should not be viewed as a failure of the proposed framework.
As discussed in Section~\ref{sec:intro}, improvements in predictive accuracy in signed networks often arise from reinforcing homophilous and antagonistic structures, which can inadvertently amplify polarization.
From this perspective, a limited accuracy reduction is a natural consequence of structural intervention aimed at polarization management. 

Importantly, larger polarization reductions do not consistently correspond
to larger drops in predictive performance. 
This suggests that the mitigation effect cannot be attributed solely by degraded embeddings or indiscriminate edge perturbations, but instead reflects targeted structural changes around antagonistic regions.
Overall, these results demonstrate that \ours\ provides a controllable mechanism for
navigating the trade-off between polarization reduction and downstream utility.


\begin{table}[t]
\centering
\caption{Impact of mitigation strength $\boldsymbol{\gamma}$ on the trade-off between polarization reduction and predictive performance (EQ3).
Larger $\boldsymbol{\gamma}$ generally yields stronger mitigation, with dataset-dependent effects on predictive performance.
}
\label{tab:eq3_gamma}
\vspace{+0.1cm}
\setlength{\tabcolsep}{6pt}
\resizebox{0.8\columnwidth}{!}{ 
\begin{tabular}{l|c|cc|c}
\hline
\textbf{Dataset} & $\boldsymbol{\gamma}$ & \textbf{Accuracy} & \textbf{Macro-F1} & $\boldsymbol{\Delta P_{\mathcal{G}',\hat{\mathbf{Z}}} (\%)}$  \\
\hline
\multirow{4}{*}{\textbf{BTC-Alpha}}
& 0.5 & 0.729 & 0.662 & -21.50 \\
& \textbf{1.0} & \textbf{0.751} & \textbf{0.691} & \textbf{-31.11} \\
& 1.5 & 0.712 & 0.662 & -35.39 \\
& 2.0 & 0.723 & 0.671 & -34.44 \\
\hline
\multirow{3}{*}{\textbf{BTC-OTC}}
& 1.0 & 0.681 & 0.590 & -22.52 \\
& \textbf{1.5} & \textbf{0.742} & \textbf{0.656} & \textbf{-30.46} \\
& 2.0 & 0.691 & 0.604 & -34.12 \\
\hline
\multirow{3}{*}{\textbf{Wiki-Elec}}
& 1.0 & 0.758 & 0.640 & -14.84 \\
& 1.5 & 0.726 & 0.629 & -25.55 \\
& \textbf{2.0} & \textbf{0.726} & \textbf{0.611} & \textbf{-27.52} \\
\hline
\end{tabular}
}
\vspace{-0.25cm}
\end{table}

\vspace{1mm}
\noindent\textbf{Results for EQ3.}
Table~\ref{tab:eq3_gamma} analyzes how the mitigation strength $\gamma$ controls the balance between polarization reduction and predictive performance.
Across all datasets, increasing $\gamma$ generally leads to stronger polarization reduction, confirming that $\gamma$ effectively governs the intensity of gray-zone augmentation.
Notably, changes in Accuracy and Macro-F1 are non-monotonic and do not exhibit systematic collapse as $\gamma$ increases.
In several cases (\eg, BTC-Alpha and BTC-OTC), stronger mitigation yields substantial polarization reduction without proportionally larger performance degradation.

These results show that $\gamma$ serves as a practical control parameter that explicitly governs the trade-off between polarization reduction and predictive utility.
Rather than relying on a single globally optimal setting, \ours\ enables network-aware mitigation that adapts intervention strength while preserving downstream performance. 

\begin{figure}[t]
    \vspace{-0.1cm}
    \centering
    \captionsetup{type=figure,skip=8pt}\includegraphics[width=\columnwidth]{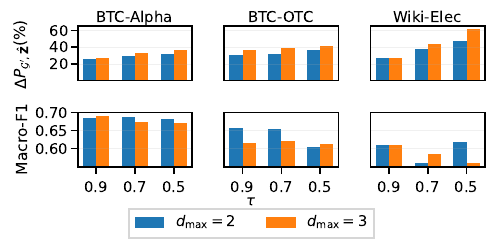}
    \vspace{-0.7cm}
    \caption{
    Sensitivity to pair-selection hyperparameters (EQ4).
    Lower $\boldsymbol{\tau}$ and larger $\boldsymbol{d_{\max}}$ yield stronger polarization reduction with predictable trade-offs in Macro-F1.
    }
    \label{fig:eq4}
    \vspace{-0.25cm}
\end{figure}

\vspace{1mm}
\noindent\textbf{Results for EQ4.}
Figure~\ref{fig:eq4} examines the sensitivity of mitigation outcomes to the pair-selection hyperparameters $\tau$ and $d_{\max}$.
As $\tau$ decreases and $d_{\max}$ increases, more polarized community pairs are selected, leading to stronger polarization reductions.
However, more aggressive selection can also increase variability or reduce predictive performance, reflecting a stronger structural intervention.
Despite differences in absolute performance, this trade-off behaves consistently across settings.
Together, these results show that $\tau$ and $d_{\max}$ act as interpretable structural controls over where and how broadly mitigation is applied, enabling systematic intervention rather than ad-hoc parameter tuning.

\vspace{1mm}
\noindent\textbf{Results for EQ5.}
We report the runtime of polarization measurement and mitigation across datasets, with
all experiments conducted on a single CPU machine equipped with an Intel Xeon Silver processor and 256 GB RAM.
Polarization measurement is consistently efficient, taking 4.9s (BTC-Alpha), 17.5s (BTC-OTC), 28.1s (Wiki-Elec), and 166.9s (Wiki-RfA).
On larger datasets such as Slashdot (6.2s) and Epinions (12.5s), measurement is even faster due to a solver-based implementation that avoids explicit Laplacian pseudoinversion (see \textbf{Appendix}).
Mitigation incurs additional cost but remains practical: it completes within tens of seconds on small and medium datasets (up to 179.0s on Wiki-RfA), and within manageable time on large-scale networks (659.0s on Slashdot and 1621.0s on Epinions).
Overall, \ours\ scales to real-world signed networks and is suitable as an offline intervention without prohibitive overhead.

\section{Related Work} \label{sec:rworks} \noindent\textbf{Signed Network Embedding.}
Early SNE methods \cite{yuan_sne_pakdd17,kim_side_www18} relied on signed random walks, while more recent GNN-based methods~\cite{derr_sgcn_icdm18,huang_sigat_icann19,huang_sdgnn_aaai21,li_signedgat_aaai20} explicitly separate positive and negative neighborhoods to model balance-theoretic structures.
Beyond representation learning, these models have been designed and evaluated in the context of specific downstream tasks such as signed link prediction~\cite{yang_friendorfrenemy_sigir12,xu_signedlatent_kdd19,fiorini_sigmanet_aaai23,huang_sdgnn_aaai21,lee_asine_sigir20,kim_trustsgcn_sigir23} and node ranking~\cite{jung2016personalized,jung2020random,lee2021look}, where embeddings serve as task-oriented representations.
More recent work directly targets polarized network settings \cite{huang_pole_wsdm22,kim_polardsn_cikm24}:
POLE \cite{huang_pole_wsdm22} enforces stronger separation between antagonistic communities,
while PolarDSN \cite{kim_polardsn_cikm24} models polarization as a temporally evolving
community-level process.
While effective at encoding polarized structures, such explicit separation can inadvertently
amplify polarization, reinforcing the escalation dynamics discussed in Section~\ref{sec:intro}.

\vspace{1mm}
\noindent\textbf{Polarization Measurement.}
A complementary line of work studies how polarization can be quantified (and mitigated), primarily  in unsigned networks \cite{hohmann_sciadv23_1,
dandekar_pnas13_1,
musco_www18_1,
garimella_wsdm17_1,
garimella_tsc18_1,
guerra_polarization_icwsm13,
coletto_controversy_osnem17,
emamgholizadeh_brw_snam20,
cota_echo_epjds19}
Under opinion dynamics frameworks, prior studies \cite{musco_www18_1,zhu_minpol_neurips21} formalize polarization as a global property of network equilibria and examine how structural
interventions can reduce it.
Notably, Musco et al.~\cite{musco_www18_1} analyze the trade-off between polarization and disagreement via optimization-based topology modification, while Hohmann et al.~\cite{hohmann_sciadv23_1} propose an effective-resistance-based measure that jointly captures opinions and network structure.
Despite these advances, existing approaches largely assume unsigned relations or explicit opinions, and do not directly address polarization measurement and mitigation in signed networks with latent representations (see Section~\ref{sec:measure}). 
\section{Conclusion} \label{sec:concl} 
In this work, we studied polarization as a structural phenomenon in signed networks that accumulates as networks evolve.
As signed network embeddings increasingly encode such polarized structures, representation learning itself can inadvertently amplify polarization in downstream analysis.
To address this issue, we proposed \ours, an embedding-aware framework that jointly measures and mitigates polarization directly in the embedding space.
\ours\ reformulates effective-resistance-based polarization measure on node embeddings and introduces a localized gray-zone-based mitigation strategy.
Experiments on real-world signed networks show that \ours\ substantially reduces polarization while preserving structural integrity, enabling controlled trade-offs with downstream utility.

\clearpage
\bibliographystyle{ACM-Reference-Format}
\bibliography{bibliography}

@inproceedings{yuan_sne_pakdd17,
  title = {SNE: Signed Network Embedding},
  author = {Yuan, Shuhan and Wu, Xintao and Xiang, Yang},
  booktitle = {Proceedings of the Pacific-Asia Conference on Knowledge Discovery and Data Mining},
  year = {2017}
}

@inproceedings{kim_side_www18,
  title = {SIDE: Representation Learning in Signed Directed Networks},
  author = {Kim, Junghwan and Park, Haekyu and Lee, Ji-Eun and Kang, U},
  booktitle = {Proceedings of the International World Wide Web Conference},
  year = {2018}
}

@inproceedings{derr_sgcn_icdm18,
  title = {Signed Graph Convolutional Networks},
  author = {Derr, Tyler and Ma, Yao and Tang, Jiliang},
  booktitle = {Proceedings of the IEEE International Conference on Data Mining},
  year = {2018}
}

@inproceedings{huang_sigat_icann19,
  title = {Signed Graph Attention Networks},
  author = {Huang, Junjie and Shen, Huawei and Hou, Liang and Cheng, Xueqi},
  booktitle = {Proceedings of the International Conference on Artificial Neural Networks},
  year = {2019}
}

@inproceedings{huang_sdgnn_aaai21,
  title = {SDGNN: Learning Node Representation for Signed Directed Networks},
  author = {Huang, Junjie and Shen, Huawei and Hou, Liang and Cheng, Xueqi},
  booktitle = {Proceedings of the AAAI Conference on Artificial Intelligence},
  year = {2021}
}

@inproceedings{huang_pole_wsdm22,
  title = {POLE: Polarized Embedding for Signed Networks},
  author = {Huang, Zexi and Silva, Arlei and Singh, Ambuj},
  booktitle = {Proceedings of the ACM International Conference on Web Search and Data Mining},
  year = {2022}
}

@inproceedings{fiorini_sigmanet_aaai23,
  title = {SigMaNet: One Laplacian to Rule Them All},
  author = {Fiorini, Stefano and Coniglio, Stefano and Ciavotta, Michele and Messina, Enza},
  booktitle = {Proceedings of the AAAI Conference on Artificial Intelligence},
  year = {2023}
}

@inproceedings{kim_polardsn_cikm24,
  title = {PolarDSN: An Inductive Approach to Learning the Evolution of Network Polarization in Dynamic Signed Networks},
  author = {Kim, Min-Jeong and Lee, Yeon-Chang and Kim, Sang-Wook},
  booktitle = {Proceedings of the ACM International Conference on Information and Knowledge Management},
  year = {2024}
}

@article{garimella_tsc18_1,
  title   = {Quantifying Controversy on Social Media},
  author  = {Garimella, Kiran and De Francisci Morales, Gianmarco and Gionis, Aristides and Mathioudakis, Michael},
  journal = {ACM Transactions on Social Computing},
  volume  = {1},
  number  = {1},
  year    = {2018}
}

@inproceedings{garimella_wsdm17_1,
  title     = {Reducing Controversy by Connecting Opposing Views},
  author    = {Garimella, Kiran and De Francisci Morales, Gianmarco and Gionis, Aristides and Mathioudakis, Michael},
  booktitle = {Proceedings of the Tenth ACM International Conference on Web Search and Data Mining},
  year      = {2017}
}

@article{hohmann_sciadv23_1,
  title   = {Quantifying Ideological Polarization on a Network Using Generalized Euclidean Distance},
  author  = {Hohmann, Marilena and Devriendt, Karel and Coscia, Michele},
  journal = {Science Advances},
  volume  = {9},
  number  = {9},
  year    = {2023}
}

@inproceedings{musco_www18_1,
  title     = {Minimizing Polarization and Disagreement in Social Networks},
  author    = {Musco, Cameron and Musco, Christopher and Tsourakakis, Charalampos E.},
  booktitle = {Proceedings of the International World Wide Web Conference},
  year      = {2018}
}

@article{dandekar_pnas13_1,
  title   = {Biased Assimilation, Homophily, and the Dynamics of Polarization},
  author  = {Dandekar, Pranav and Goel, Ashish and Lee, David T.},
  journal = {Proceedings of the National Academy of Sciences},
  volume  = {110},
  number  = {15},
  year    = {2013}
}

@article{cartwright_psychrev56_1,
  title   = {Structural Balance: A Generalization of Heider's Theory},
  author  = {Cartwright, Dorwin and Harary, Frank},
  journal = {Psychological Review},
  volume  = {63},
  number  = {5},
  year    = {1956}
}

@article{heider_jpsych46_1,
  title   = {Attitudes and Cognitive Organization},
  author  = {Heider, Fritz},
  journal = {The Journal of Psychology},
  volume  = {21},
  number  = {1},
  year    = {1946}
}

@inproceedings{leskovec_chi10_1,
  title     = {Signed Networks in Social Media},
  author    = {Leskovec, Jure and Huttenlocher, Daniel and Kleinberg, Jon},
  booktitle = {Proceedings of the SIGCHI Conference on Human Factors in Computing Systems},
  year      = {2010}
}

@inproceedings{bonchi_cikm19_1,
  title     = {Discovering Polarized Communities in Signed Networks},
  author    = {Bonchi, Francesco and Galimberti, Edoardo and Gionis, Aristides and Ordozgoiti, Bruno and Ruffo, Giancarlo},
  booktitle = {Proceedings of the 28th ACM International Conference on Information and Knowledge Management},
  year      = {2019}
}

@inproceedings{guha_www04_1,
  title     = {Propagation of Trust and Distrust},
  author    = {Guha, R. and Kumar, R. and Raghavan, P. and Tomkins, A.},
  booktitle = {Proceedings of the International World Wide Web Conference},
  year      = {2004}
}

@inproceedings{massa_recsys07_1,
  title     = {Trust-aware recommender systems},
  author    = {Massa, Paolo and Avesani, Paolo},
  booktitle = {Proceedings of the ACM Conference on Recommender Systems},
  year      = {2007}
}

@inproceedings{kunegis_www09_1,
  title     = {The Slashdot Zoo: Mining a Social Network with Negative Edges},
  author    = {Kunegis, J. and Lommatzsch, A. and Bauckhage, C.},
  booktitle = {Proceedings of the International World Wide Web Conference},
  year      = {2009}
}

@article{jung2020random,
  title   = {Random Walk-Based Ranking in Signed Social Networks: Model and Algorithms},
  author  = {Jung, Jinhong and Jin, Woojeong and Kang, U},
  journal = {Knowledge and Information Systems},
  volume  = {62},
  number  = {2},
  year    = {2020}
}

@inproceedings{jung2016personalized,
  title     = {Personalized Ranking in Signed Networks Using Signed Random Walk with Restart},
  author    = {Jung, Jinhong and Jin, Woojeong and Sael, Lee and Kang, U},
  booktitle = {Proceedings of the IEEE International Conference on Data Mining},
  year      = {2016}
}

@inproceedings{lee2021look,
  title     = {Look Before You Leap: Confirming Edge Signs in Random Walk with Restart for Personalized Node Ranking in Signed Networks},
  author    = {Lee, Wonchang and Lee, Yeon-Chang and Lee, Dongwon and Kim, Sang-Wook},
  booktitle = {Proceedings of the ACM SIGIR Conference on Research and Development in Information Retrieval},
  year      = {2021}
}

@inproceedings{haddadan2021repbublik,
  title     = {RePBubLik: Reducing Polarized Bubble Radius with Link Insertions},
  author    = {Haddadan, Shahrzad and Menghini, Cristina and Riondato, Matteo and Upfal, Eli},
  booktitle = {Proceedings of the ACM International Conference on Web Search and Data Mining},
  year      = {2021}
}

@article{klein_randic93_resistance,
  title   = {Resistance Distance},
  author  = {Klein, Douglas J. and Randi{\'c}, Milan},
  journal = {Journal of Mathematical Chemistry},
  volume  = {12},
  number  = {1},
  year    = {1993}
}

@inproceedings{coscia20_generalized_euclidean,
  title     = {Generalized Euclidean Measure to Estimate Network Distances},
  author    = {Coscia, Michele},
  booktitle = {Proceedings of the International AAAI Conference on Web and Social Media},
  year      = {2020}
}

@article{delvalle2018_echo,
  title   = {Echo Chambers in Parliamentary Twitter Networks: The Catalan Case},
  author  = {Del Valle, Marc Esteve and Borge Bravo, Rosa},
  journal = {International Journal of Communication},
  volume  = {12},
  year    = {2018}
}

@article{delvalle2022_political_interaction,
  title   = {Political Interaction Beyond Party Lines: Communication Ties and Party Polarization in Parliamentary Twitter Networks},
  author  = {Esteve Del Valle, Marc and Broersma, Marcel and Ponsioen, Arnout},
  journal = {Social Science Computer Review},
  volume  = {40},
  number  = {3},
  year    = {2022}
}

@article{gomez2009_signed_community,
  title   = {Analysis of Community Structure in Networks of Correlated Data},
  author  = {G{\'o}mez, Sergio and Jensen, Pablo and Arenas, Alex},
  journal = {Physical Review E},
  volume  = {80},
  number  = {1},
  year    = {2009}
}

@article{jolliffe1990_pca,
  title   = {Principal Component Analysis: A Beginner's Guide—I. Introduction and Application},
  author  = {Jolliffe, Ian T.},
  journal = {Weather},
  volume  = {45},
  number  = {10},
  year    = {1990}
}

@inproceedings{macqueen1967_kmeans,
  title     = {Some Methods for Classification and Analysis of Multivariate Observations},
  author    = {McQueen, James B.},
  booktitle = {Proceedings of the Berkeley Symposium on Mathematical Statistics and Probability},
  year      = {1967}
}

@inproceedings{sharma2023_dynamic_signed,
  title     = {Representation Learning in Continuous-Time Dynamic Signed Networks},
  author    = {Sharma, Kartik and Raghavendra, Mohit and Lee, Yeon-Chang and M., Anand Kumar and Kumar, Srijan},
  booktitle = {Proceedings of the ACM International Conference on Information and Knowledge Management},
  year      = {2023}
}

@article{fritz2025,
  author       = {Cornelius Fritz and
                  Marius Mehrl and
                  Paul W. Thurner and
                  G{\"{o}}ran Kauermann},
  title        = {Exponential Random Graph Models for Dynamic Signed Networks: An Application to International Relations},
  journal      = {Political Analysis},
  volume       = {33},
  year         = {2025}
}

@article{MorrisonKG23,
  author       = {Megan Morrison and
                  J. Nathan Kutz and
                  Michael Gabbay},
  title        = {Transitions between peace and systemic war as bifurcations in a signed
                  network dynamical system},
  journal      = {Netw. Sci.},
  volume       = {11},
  number       = {3},
  year         = {2023}
}

@article{DiazDiazBE24,
  author       = {Fernando Diaz{-}Diaz and
                  Paolo Bartesaghi and
                  Ernesto Estrada},
  title        = {Mathematical modeling of local balance in signed networks and its
                  applications to global international analysis},
  journal      = {J. Appl. Math. Comput.},
  volume       = {70},
  number       = {6},
  year         = {2024}
}

@inproceedings{lee_asine_sigir20,
  title   = {ASiNE: Adversarial Signed Network Embedding},
  author  = {Lee, Yeon-Chang and Seo, Nayoun and Han, Kyungsik and Kim, Sang-Wook},
  booktitle = {Proceedings of the ACM SIGIR Conference on Research and Development in Information Retrieval},
  year    = {2020}
}

@inproceedings{kim_trustsgcn_sigir23,
  title   = {TrustSGCN: Learning Trustworthiness on Edge Signs for Effective Signed Graph Convolutional Networks},
  author  = {Kim, Min-Jeong and Lee, Yeon-Chang and Kim, Sang-Wook},
  booktitle = {Proceedings of the ACM SIGIR Conference on Research and Development in Information Retrieval},
  year    = {2023}
}

@article{guerra_polarization_icwsm13,
  title   = {A Measure of Polarization on Social Media Networks Based on Community Boundaries},
  author  = {Guerra, Pedro and Meira Jr., Wagner and Cardie, Claire and Kleinberg, Robert},
  journal = {Proceedings of the International AAAI Conference on Web and Social Media},
  year    = {2013},
  volume  = {7},
  number  = {1}
}

@article{garcia_polarization_policyinternet15,
  title   = {Ideological and Temporal Components of Network Polarization in Online Political Participatory Media},
  author  = {Garcia, David and Abisheva, Adiya and Schweighofer, Simon and Serd{\"u}lt, Uwe and Schweitzer, Frank},
  journal = {Policy \& Internet},
  year    = {2015},
  volume  = {7},
  number  = {1}
}

@article{cinelli_echo_pnas21,
  title   = {The Echo Chamber Effect on Social Media},
  author  = {Cinelli, Matteo and De Francisci Morales, Gianmarco and Galeazzi, Alessandro and Quattrociocchi, Walter and Starnini, Michele},
  journal = {Proceedings of the National Academy of Sciences},
  year    = {2021},
  volume  = {118},
  number  = {9}
}

@article{santos_linkrec_pnas21,
  title   = {Link Recommendation Algorithms and Dynamics of Polarization in Online Social Networks},
  author  = {Santos, Fernando P. and Lelkes, Yphtach and Levin, Simon A.},
  journal = {Proceedings of the National Academy of Sciences},
  year    = {2021},
  volume  = {118},
  number  = {50}
}

@article{friedkin_johnsen_opinion90,
  title   = {Social Influence and Opinions},
  author  = {Friedkin, Noah E. and Johnsen, Eugene C.},
  journal = {The Journal of Mathematical Sociology},
  year    = {1990},
  volume  = {15},
  number  = {3--4}
}

@inproceedings{li_signedgat_aaai20,
  title     = {Learning Signed Network Embedding via Graph Attention},
  author    = {Li, Yu and Tian, Yuan and Zhang, Jiawei and Chang, Yi},
  booktitle = {Proceedings of the AAAI Conference on Artificial Intelligence},
  year      = {2020}
}

@inproceedings{yang_friendorfrenemy_sigir12,
  title     = {Friend or Frenemy? Predicting Signed Ties in Social Networks},
  author    = {Yang, Shuang-Hong and Smola, Alexander J. and Long, Bo and Zha, Hongyuan and Chang, Yi},
  booktitle = {Proceedings of the ACM SIGIR Conference on Research and Development in Information Retrieval},
  year      = {2012}
}

@inproceedings{xu_signedlatent_kdd19,
  title     = {Link Prediction with Signed Latent Factors in Signed Social Networks},
  author    = {Xu, Pinghua and Hu, Wenbin and Wu, Jia and Du, Bo},
  booktitle = {Proceedings of the ACM SIGKDD International Conference on Knowledge Discovery and Data Mining},
  year      = {2019}
}

@article{coletto_controversy_osnem17,
  title   = {Automatic Controversy Detection in Social Media: A Content-Independent Motif-Based Approach},
  author  = {Coletto, Mauro and Garimella, Kiran and Gionis, Aristides and Lucchese, Claudio},
  journal = {Online Social Networks and Media},
  year    = {2017},
  volume  = {3--4}
}

@article{emamgholizadeh_brw_snam20,
  title   = {A Framework for Quantifying Controversy of Social Network Debates Using Attributed Networks: Biased Random Walk},
  author  = {Emamgholizadeh, Hanif and Nourizade, Milad and Tajbakhsh, Mir Saman and Hashminezhad, Mahdieh and Esfahani, Farzaneh Nasr},
  journal = {Social Network Analysis and Mining},
  year    = {2020},
  volume  = {10},
  number  = {1}
}

@article{cota_echo_epjds19,
  title   = {Quantifying Echo Chamber Effects in Information Spreading over Political Communication Networks},
  author  = {Cota, Wesley and Ferreira, Silvio C. and Pastor-Satorras, Romualdo and Starnini, Michele},
  journal = {EPJ Data Science},
  year    = {2019},
  volume  = {8},
  number  = {1}
}

@inproceedings{zhu_minpol_neurips21,
  title     = {Minimizing Polarization and Disagreement in Social Networks via Link Recommendation},
  author    = {Zhu, Liwang and Bao, Qi and Zhang, Zhongzhi},
  booktitle = {Advances in Neural Information Processing Systems},
  year      = {2021}
}

\clearpage
\appendix
\appendix
\section{Appendix}~\label{app}

\noindent In this appendix, we provide additional implementation details that complement the experimental setup described in the main paper.

\subsection{Further Implementation Details}~\label{app:setup}

\vspace{+1mm}
\noindent\textbf{Graph preprocessing and community construction.}
The number of communities $k$ is estimated on the training graph
using signed Louvain clustering.
We run the algorithm 10 times with different random seeds and record
the number of detected communities after discarding
small communities.
Communities with fewer than 30 nodes are treated as noise and excluded;
for large-scale datasets (Slashdot and Epinions), this threshold is
increased to 500.
The final value of $k$ is determined as the mode across runs.

For Slashdot and Epinions, we further apply pruning during polarized community search to reduce computational cost.
We retain only communities with at least 1000 nodes and compute a lightweight pruning score for each remaining community pair based on the Euclidean distance between community centroids in the aligned embedding space.
Only the top 50 most distant pairs are retained for subsequent polarized community scoring.

\vspace{+1mm}
\noindent\textbf{Embedding model, training, and evaluation protocol.}
Node embeddings are learned using SGCN, with hyperparameters tuned via grid search on the validation set based on Macro-F1.
The search space includes embedding dimensions
$\{32, 64, 128\}$, numbers of layers $\{2, 3, 4\}$, and learning rates
$\{0.05, 0.01, 0.005, 0.001, 0.0005\}$.
Models are trained for up to 400 epochs with early stopping based on
the validation loss (patience = 10).

Signed link prediction is formulated as a three-class classification
task over positive edges, negative edges, and non-edges.
For each split, non-edges are sampled to match the total number of observed signed edges and fixed per random seed.
When mitigation modifies the training graph, validation and test splits, along with their corresponding non-edge samples, are kept unchanged to ensure fair comparison.

\vspace{+1mm}
\noindent\textbf{Polarization computation and scalability.}
Naively computing the effective-resistance-based polarization
requires forming the Laplacian pseudoinverse $L^{\dagger}$,
which has cubic time complexity $\mathcal{O}(|V|^{3})$.
Instead, we compute quadratic forms $b^{\top}L^{\dagger}b$ by solving sparse linear systems using a Laplacian solver.
This reduces the computational cost to near-linear or mildly super-linear time in the number of nonzero entries of the Laplacian entries.
This solver-based computation is applied both for network-level polarization measurement and for repeated pairwise evaluations during polarized community search, enabling scalability to large-scale datasets.

\vspace{+1mm}
\noindent\textbf{Mitigation hyperparameters and validation protocol.}
Mitigation-related hyperparameters are tuned on the validation set.
For BTC-Alpha, BTC-OTC, Wiki-Elec, and Wiki-RfA, we vary the pair-selection threshold
$\tau \in \{0.5, 0.6, 0.7, 0.8, 0.9\}$, the maximum selection count
$d_{\max} \in \{2, 3\}$, and the mitigation strength
$\gamma \in \{1.0, 1.5, 2.0\}$, while fixing the negative edge scale
factor to $\eta = 0.1$.
For large-scale datasets (Slashdot and Epinions), we fix $\gamma=1$
and vary $(\tau, d_{\max})$ over $\tau \in \{0.2, 0.3\}$ and
$d_{\max} \in \{3, 4\}$.
All results are averaged over five random seeds, with selected configurations reported in Table~\ref{tab:mitigation_hyper_results}.

\begin{table}[t]
\centering
\caption{Selected mitigation hyperparameters and their effects on  performance and polarization
All $\Delta$ values are reported in \% and averaged over five random seeds.}
\label{tab:mitigation_hyper_results}
\begin{tabular}{l|ccc|cc|c}
\toprule
Dataset 
& $\tau$ & $d_{\max}$ & $\gamma$ 
& $\Delta$Acc 
& $\Delta$F1 
& $\Delta \boldsymbol{P_{\mathcal{G},\hat{\mathbf{Z}}}}$ \\
\midrule
BTC-Alpha  & 0.6 & 2 & 1.0 & +0.84 & -0.90 & -31.11 \\
BTC-OTC    & 0.9 & 2 & 1.5 & -6.76 & -8.17 & -30.46 \\
Wiki-Elec  & 0.9 & 2 & 2.0 & -5.86 & -9.25 & -27.52 \\
Wiki-RfA   & 0.7 & 2 & 1.0 & -2.21 & -0.75 & -36.96 \\
Slashdot   & 0.2 & 4 & 1.0 & -0.06 & -0.32 & -13.03 \\
Epinions   & 0.2 & 4 & 1.0 & -0.58 & -0.09 & -22.28 \\
\bottomrule
\end{tabular}
\end{table}

\vspace{+1mm}
\noindent\textbf{Computing environment.}
All experiments were implemented in Python 3.10.
SGCN training was performed on NVIDIA RTX 4090 and RTX 5090 GPUs.
Polarization measurement and mitigation components, including polarized community search and solver-based computations, were executed on a single CPU machine equipped with an Intel Xeon Silver processor and 256 GB RAM.

\end{document}